\documentclass[12pt,onecolumn]{article}	
\usepackage[top=1in, bottom=1in, left=1in, right=1in]{geometry}
\usepackage{graphicx}
\usepackage{amssymb}
\usepackage{amsmath}  				
\usepackage{epstopdf}  				
\usepackage{graphicx,float,wrapfig} 	
\usepackage{color}					
\usepackage[colorlinks,linktocpage=true]{hyperref}		
\usepackage[font=small,labelfont=bf,textfont=sf,
labelsep=quad]{caption}				

\usepackage[superscript,space]{cite}	

\hypersetup{						
  citecolor= black,
  linkcolor=black,
}

\usepackage{setspace}   

\usepackage [english]{babel}
\usepackage [english = american]{csquotes}
\MakeOuterQuote{"}

\doublespace                   				

\usepackage[protrusion=false,expansion=false]{microtype}

\let\baraccent=\= 
\renewcommand{\=}[1]{\stackrel{#1}{=}} 

\begin{document}

\title{Vortex arrays and ciliary tangles underlie the feeding--swimming tradeoff in starfish larvae}

\author
{William Gilpin$^{1}$, Vivek N. Prakash$^{2}$, Manu Prakash$^{2 \ast}$\\
\\
\normalsize{$^{1}$Department of Applied Physics, $^{2}$Department of Bioengineering, }\\
\normalsize{Stanford University, Stanford, CA}\\
\\
\normalsize{$^\ast$To whom correspondence should be addressed; E-mail:  manup@stanford.edu}
\\
}


\maketitle

\newpage

\begin{abstract}
Many marine invertebrates have larval stages covered in linear arrays of beating cilia, which propel the animal while simultaneously entraining planktonic prey.\cite{strathmann1971feeding} These bands are strongly conserved across taxa spanning four major superphyla,\cite{nielsen1987structure,strathmann1993hypotheses} and they are responsible for the unusual morphologies of many invertebrate larvae.\cite{lacalli1990ciliary,von2013pilidium} However, few studies have investigated their underlying hydrodynamics.\cite{strathmann1975larval,hart1991particle} Here, we study the ciliary bands of starfish larvae, and discover a beautiful pattern of slowly-evolving vortices that surrounds the swimming animals. Closer inspection of the bands reveals unusual ciliary "tangles" analogous to topological defects that break-up and re-form as the animal adjusts its swimming stroke. Quantitative experiments and modeling demonstrate that these vortices create a physical tradeoff between feeding and swimming in heterogenous environments, which manifests as distinct flow patterns or "eigenstrokes" representing each behavior---potentially implicating neuronal control of cilia. This quantitative interplay between larval form and hydrodynamic function may generalize to other invertebrates with ciliary bands, and illustrates the potential effects of active boundary conditions in other biological and synthetic systems.
\end{abstract}

\newpage

Many marine invertebrates develop in two stages, growing from swimming larvae to bottom-dwelling adults. This developmental dichotomy mandates physical tradeoffs, resulting in a stunning variety of larval morphologies that have transfixed biologists and artists alike for more than a century\cite{agassiz1877north}. Common to many larval forms are dense ciliary bands encircling the body, which propel the animal during escape and dispersal while also entraining motile plankton, their primary food source\cite{strathmann1971feeding}. Ciliary bands are linear arrays of densely-packed cilia, which form morphologies that are as diverse as the remarkable range of organisms in which they appear (which spans four major superphyla)\cite{strathmann1993hypotheses,amemiya2015development}. Although several studies have emphasized the importance of ciliary bands in phylogenetic reconstruction\cite{nielsen1987structure,amemiya2015development}, few have quantitatively investigated them, particularly the physical mechanism by which they simultaneously maximize swimming flow while precisely capturing food particles---an intrinsic link between larval morphology and behavior. 

Here, we observe that the larva of the common starfish {\it Patiria miniata} generates around its body a beautiful and intricate array of counter-rotating vortices, which merge and split in a complex manner over time (Fig. 1A, Supplementary Video 1; 29 larvae). We find these patterns in both freely-swimming and animals held stationary, and the larvae swim irregularly as these currents evolve in time (Supplementary Videos 2, 3, 4; 37 larvae). Slowly-varying vortices have been noted in several marine larvae, leading some authors to speculate that these structures play a crucial role in resolving the competing requirements of swimming and feeding\cite{strathmann1971feeding,hart1991particle,lacalli1990ciliary,von2013pilidium,strathmann2006good,burke1983structure}.  Here we investigate this conjecture using a combination of live imaging and mathematical modeling, and we reveal a subtle interplay between organism-level band morphology and microscopic ciliary "tangles," which jointly produce the distinctive, dynamic flow fields seen around the organism. We further analyze these flow fields in order to identify separate hydrodynamic "eigenstrokes" corresponding to distinct feeding and swimming behaviors. To our knowledge, our results provide the first quantitative study of the relationship between form and function in ciliary bands, providing a biophysical example of how the competing hydrodynamics of feeding and swimming may unify morphology and behavior in marine invertebrate development \cite{strathmann2006good}.

Ciliary bands in a {\it Patiria miniata} larva surround the animal's periphery, dividing the body into pre- and post-oral hoods and the circumoral field. Swimming occurs with the pre-oral hood facing upward and anterior facing forward\cite{strathmann1971feeding} (Fig. 1B). The circumoral bands are primarily responsible for the observed vortices, which localize to the animal's midplane. The vortices remain mostly symmetric across the body axis (although this symmetry varies over time), and minor left-right differences in body morphology create small asymmetries in the vortex locations. The largest vortices occur at the posterior end of the larva, while smaller vortices localize to dense portions of the band near the side lobes. Stagnation points with zero net flow form where two counter-rotating vortices intersect. Consistent with previous larval feeding experiments,\cite{strathmann1975larval,hart1991particle} algae that pass through the vortex field and contact the body surface are captured and then transported either along the ciliary band or by surface currents in the circumoral field until they reach the oral hood, where they aggregate before being ingested. An electron micrograph of a larva reveals shorter cilia covering the body surface that facilitate this transfer, with higher line densities of cilia ($\sim 3$ cilia/$\mu$m) occurring along the ciliary band than on the body surface ($< 0.1$ cilia/$\mu$m) (Fig. 1C) \cite{strathmann1971feeding,hart1996variation}. This unusually high density and configuration of cilia enables physical interactions between the cilia of neighboring cells, which is uncommon in single cell ciliates \cite{goldstein2015green,guasto2012fluid}.

\begin{figure}
{
\centering
\includegraphics[width=\linewidth]{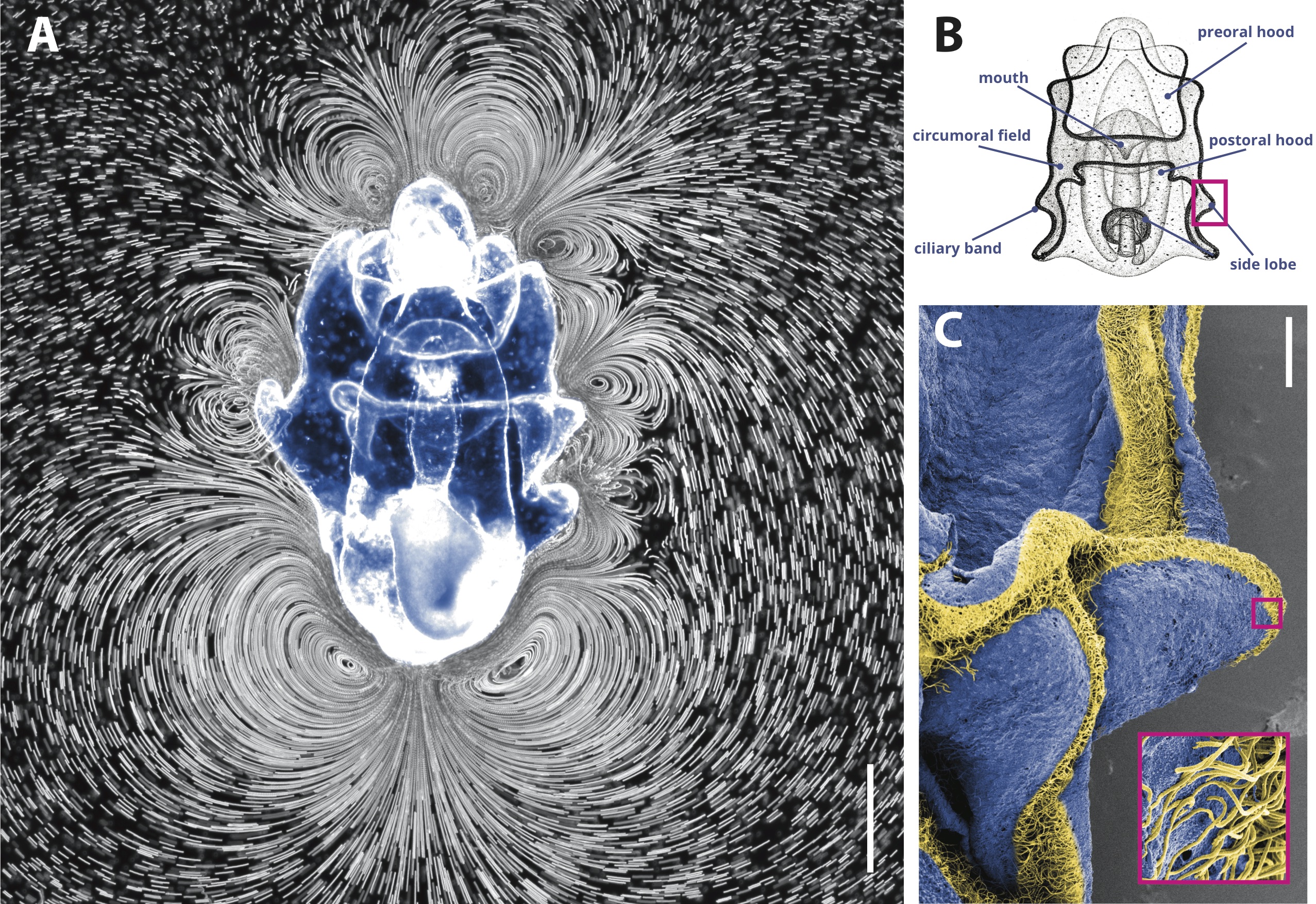}
\caption{
{\bf Bat star larvae generate complex arrays of vortices.} ({\bf A}) An eight-week-old starfish larva feeds while held in a $\sim 0.5$ mm gap between a glass slide and coverslip. Immobilization enables visualization of 9 counter-rotating vortices by stacking a 3 s dark-field video of $6$ $\mu$m tracer particles \cite{gilpin2016flowtrace}. Swimming direction upwards, scale bar: 350 $\mu$m. ({\bf B}) The frontal plane anatomy of a starfish larva, with the ciliary band highlighted as a dark line (adapted from a lithograph by A. Agassiz, 1877). ({\bf C}) A false-color SEM highlighting the ciliary band (yellow). Scale bar: 20 $\mu$m.}
\label{fig1vortices}
}
\end{figure}  

Closer inspection of the cilia reveals a typical stroke pattern (Fig. 2D, Supplementary Video 14) with a mean beat frequency of 6 Hz and length of 30 $\mu$m, consistent with previous measurements\cite{strathmann1971feeding}. These cilia create flow speeds of up to 300 $\mu$m/s tangential to the surface (and negligible flow through the surface), but which vary in magnitude and direction across the organism---giving rise to the observed vortices. Surprisingly, we observe unusual arrangements of the cilia wherever the surface beat direction reverses (Fig. 2), which we term "defects" by analogy to topological defects observed in nematic liquid crystals and magnetic spins\cite{chaikin2000principles}. Cilia beating towards (away from) each other create outwards (inwards) flows and a characteristic tangle (splay) defect, hereafter denoted by + (--) (Supplementary Videos 5, 13; 25 larvae). The near-field streamlines match the predicted flow along a wall with a prescribed tangential flow velocity that undergoes a sudden change in sign (Fig. 2B). We image the defects and flow fields simultaneously, and we observe that each defect creates a boundary between two vortices of opposite rotation direction (Fig. 2C, Supplementary Video 6). Consistent with the time-varying nature of larval swimming, defects can be seen to break up and re-form in place over timescales $\sim\!10$ s (Fig. 2E, Supplementary Videos 7, 8), creating corresponding changes in the vortices. The defect locations appear to be determined anatomically, and typical locations for a larva are indicated in Figure 2F.

\begin{figure}
{
\centering
\includegraphics[width=.8\linewidth]{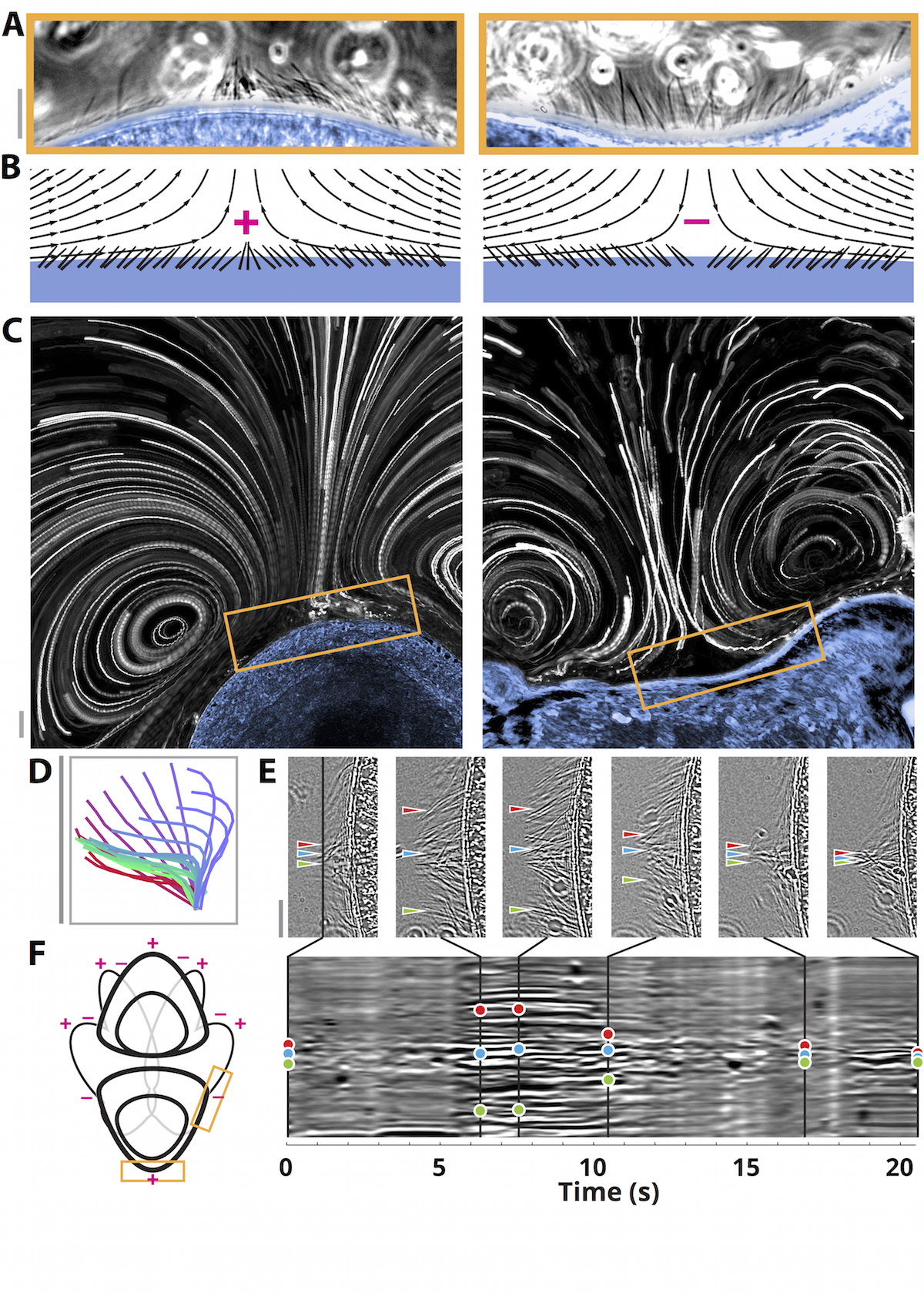}
\caption{
{\bf Transient ciliary defects underly vortex formation.} ({\bf A}) Boundaries between vortices require local reversals in the surface beat direction of cilia, which create tangles (left, +) or splays (right, --) depending on the rotation directions of the vortices. ({\bf B}) Streamlines for a model flow generated by a sign change in the tangential velocity imposed on a surface, with the orientation of the cilia indicated schematically. ({\bf C}) Zoomed out view of the same video shown in ({\bf A}), with frames stacked over 4.5 s to show pathlines of advected particles near the defect. ({\bf D}) A trace of a typical ciliary stroke near a defect, color code indicates evenly-spaced timepoints (green to red ) ({\bf E}) A kymograph of a tangle defect (with the locations of three representative cilia marked) with six video frames called out. ({\bf F}) The locations of + and -- defects on a typical larvae, with the locations in the previous panels indicated. Scale bars: 30 $\mu$m.
}
\label{cilia}
}
\end{figure}

When motile algae ({\it Rhodomonas lens}, $\sim 6$ $\mu$m) are added to seawater containing $0.75$ $\mu$m beads, they appear to swim freely until they enter a high-speed ($\sim 100$ $\mu$m/s) region near the body, where they are entrained and repeatedly swept towards the animal's surface by surrounding vortices (Fig. 3A, Supplementary Video 9). Upon approach to the body they may be captured through instantaneous reverse beats of single cilia (Fig. 3B, Supplementary Video 10). This direct-capture mechanism has been reported in other marine plankton---including starfish---as a manifestation of "interception feeding" \cite{kiorboe2011zooplankton, strathmann1975larval,von2013pilidium}. Hydrodynamically, vortex-mediated surface capture occurs due to short portions of the ciliary band beating in reverse relative to the direction of the prevailing swimming current \cite{lacalli1990ciliary,strathmann1971feeding,hart1991particle,mackie1969electrical}, which draw particles close to the surface for capture. These regions of the ciliary band span the region between two defects of opposite sign (Fig. 3C). The local structure of this mechanism is consistent across multiple animals; velocity line traces through the vortices overlap when rescaled by vortex speed and size (Fig. 3D), consistent with the power-law dropoff expected for a confined active swimmer\cite{magar2003nutrient,pepper2009nearby}. 

We incorporate these observations into a simple mathematical model that captures the tradeoff between near-field feeding and the far-field swimming currents, which are both fully determined by the boundary condition imposed at the larvae's surface. The flows are dominated by viscous rather than inertial forces ($Re < 0.003$), and so we adapt the classic "squirmer" model for the propulsive currents generated by a two-dimensional, circular swimmer imposing an arbitrary tangential velocity profile along its surface---here representing the ciliary band outlining the frontal plane\cite{blake1971self}. The velocity field around the swimmer is given by
\begin{align}
v_r(r,\theta) &=  \dfrac12 B_1 \cos\theta \left(  \left(\dfrac{a}{r} \right)^2 - 1  \right) + \sum_{n=2}^{\infty} \dfrac{n}{2} B_n \cos(n \theta)  \left(  \left(	\dfrac{a}{r} \right)^{n+1} -\left( \dfrac{a}{r} \right)^{n-1}	 \right) \notag\\
v_\theta(r, \theta) &= \dfrac12 B_1 \sin\theta \left(  \left(\dfrac{a}{r} \right)^2  + 1  \right) + \sum_{n=2}^{\infty} \dfrac{1}{2} B_n \sin(n \theta)  \left(  n\left(\dfrac{a}{r} \right)^{n+1} -(n-2)\left( \dfrac{a}{r} \right)^{n-1}	 \right)
\label{squirmer}
\end{align}
where $v_r,v_\theta$ denote the radial and azimuthal velocity components in the co-moving frame. The swimming speed is determined solely by the first-order mode, $v_{swim} = -(1/2)B_1$, because all other terms dependent on $r$ vanish as $r\rightarrow \infty$. The squirmer model has been successfully applied to single-celled organisms with localized thrust-generating appendages, which create "pusher/puller" currents that manifest as a single pair of vortices at the fore/aft of the animal.\cite{goldstein2015green,guasto2012fluid} To our knowledge, ours is the first organism in which more than two such vortices are present, and so we introduce higher-order "squirming" terms ($B_n>0; n > 2$)---which we attribute to the more complex neurociliary control found in invertebrate larvae\cite{strathmann1971feeding,lacalli1990ciliary}.

We restrict ourselves to a set of "two-mode" swimmers, each comprising a superposition of a swimming mode ($B_1$) that generates propulsion via a streaming flow past the surface, and an "$n^{th}$ order" mode ($B_n, n \geq 2$) that has $2n$ sign changes (ciliary defects) along the surface that generate reversal regions (forming $2 n - 2$ vortices for $n<5$, $2 n$ vortices for $n\geq 5$). Higher values of $n$ result in smaller average distances between defects of opposite sign. Squirmers with different $n$ are normalized by the maximum surface cilia beat speed and the size of the vortex region (the latter constraint is a geometric rather than biological; see Supplementary Methods). Streamlines for $n = 1,3,4$ are shown in Figure 3E, and the corresponding swimming speeds are observed to decrease with $n$ (Fig. 3F, violet trace).

We introduce two-mode swimmers as a minimal model of locomotion with vorticity in order to quantify the effect of vortices on the feeding ability. In invertebrate zoology, the feeding rate (food particles per unit time) is typically estimated using the "clearance rate," or the volume of water searched by the animal per unit time\cite{hart1991particle,kiorboe2011zooplankton,rothschild1988small,rubenstein1977mechanisms}. This quantity has been investigated for a variety of marine larvae, and across many taxa it is generally found to increase with both the total ciliary band length and larval size \cite{hart1996variation,strathmann1971feeding}. Here we represent the total feeding rate $\dot N$ as the product of three factors,
\begin{equation}
\dot N = \rho \,\sigma\, v_{swim}
\label{rate}
\end{equation}
where $\rho$ is the nutrient density, $\sigma$ is the hydrodynamic capture cross section (which may be larger than the animal's body cross section), and $v_{swim}$ is the swimming speed (which determines how often the animal encounters new food particles). The product $\sigma v_{swim}$ is a volume flux that is equivalent to the clearance rate \cite{kiorboe2011zooplankton,marrase1990grazing,rothschild1988small,rubenstein1977mechanisms}. In previous clearance rate experiments conducted in homogenous environments\cite{strathmann1971feeding,hart1991particle}, the feeding rate is maximized when the swimming speed $v_{swim}$ is maximized, as was found in recent theoretical work on osmotrophic feeders\cite{michelin2011optimal,magar2003nutrient}. However, many zooplankton survive in extremely heterogenous environments, in which their rate of feeding is not limited by their swimming speed but rather by their ability to 1) encounter new microalgal patches, and 2) efficiently trap and capture food upon encounter \cite{fenchel2002microbial,siegel1998resource,jiang2002chemoreception,marrase1990grazing}. In these circumstances, heterogeneity may encourage "cross sectional feeding" (maximizing $\sigma$) instead of "flux feeding" (maximizing $\rho\,v_{swim}$). This description matches findings for other model swimmers with feeding\cite{tam2011optimal}, and it supports prior experimental observations\cite{strathmann1971feeding,hart1991particle} that longer ciliary bands create a larger feeding rates even when they do not provide an increase in swimming speed\cite{hart1996variation}.

We estimate $\sigma$ numerically by simulating the trajectory of a line source of particles incident on the squirmer, and probabilistically capture particles that pass within a fixed "interception distance" of the surface (Supplementary Methods). For passive particles (as used in the simulations here), this is equivalent to determining the fraction of incoming streamlines that pass within the interception distance, weighted by their residence time. Example trajectories are shown in Figure 3G, with those of captured particles highlighted in red. The number of captured particles is rescaled by the initial number of particles, the squirmer diameter, and the interception distance, in order to produce a non-dimensional cross-section which increases with increasing number of vortices (Fig. 3F). This increase arises primarily geometrically due to more particles being drawn into the interception region via streamline compression, as well as due to particles entering the interception region repeatedly and for longer times when more vortices are present---observations consistent with prior larval feeding experiments \cite{hart1991particle}. Adding random noise to the simulated particle trajectories in order to emulate algal diffusivity or activity further increases the capture cross section (Supplementary Analysis).

The scaling of $\sigma$ with $n$ suggests a plausible reason for the striking flow fields observed around starfish larvae, for which additional vortices increase feeding cross section at the expense of swimming speed. This tradeoff is also consistent with the constraints of a fixed energy budget, since the swimmer either dissipates energy due to surface drag from swimming, or vortex core drag from feeding \cite{blake1971self,guasto2012fluid}. Our results also support previous evidence that organisms with finite interception/capture rates should favor vortical currents that draw particles near their surface\cite{kiorboe2011zooplankton,jiang2002chemoreception}.

\begin{figure}
{
\centering
\includegraphics[width=\linewidth]{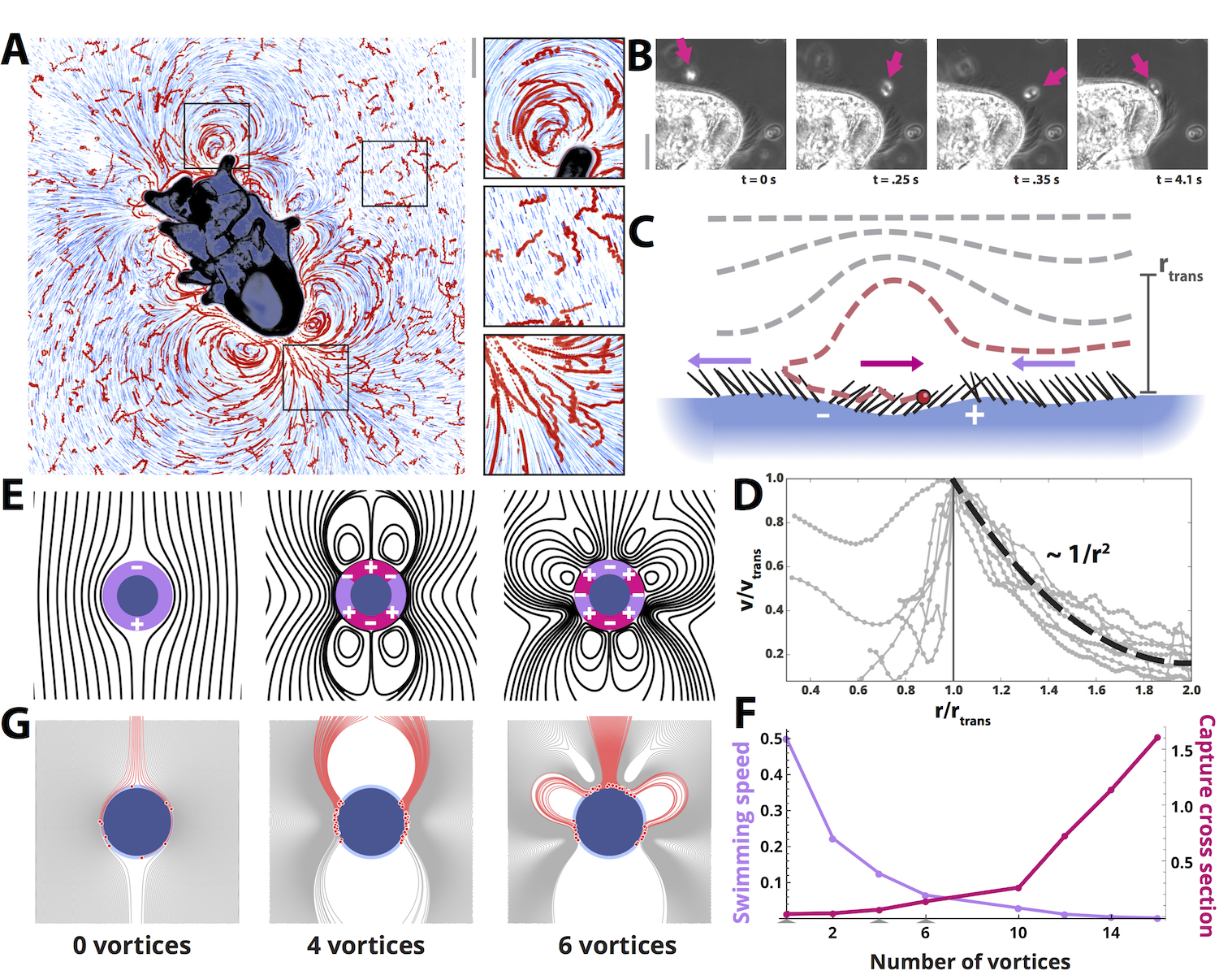}
\caption{
{\bf Vorticity arises due to a tradeoff between feeding and swimming currents.} ({\bf A}) The flow field generated by a  larvae (blue) in a mixture of passive beads (light blue) and active algae (red). Video frames stacked over 60 s, scale bar: 300 $\mu$m. ({\bf B}) A typical algae capture sequence, with the algae orientation indicated with an arrow. Scale bar: 40 $\mu$m. ({\bf C}) A schematic of local ciliary reversal between two defects capturing a particle, which is then passed along the surface by cilia. ({\bf D})  Line traces of the velocity component parallel to the body surface, taken through the largest vortex for 8 organisms (a parabola is overlaid on the decay portions). Streamlines ({\bf E}) and trajectories of incident algae particles ({\bf G}) for the "two-mode" squirmer for different orders $n$. Swimming (purple) and reversal (magenta) ciliary regions are indicated, and captured particle trajectories are highlighted in red. ({\bf F}) The swimming speed (purple) and capture cross section (magenta) as a function of the number of vortices}
\label{modes}
}
\end{figure}


\begin{figure}
{
\centering
\includegraphics[width=.8\linewidth]{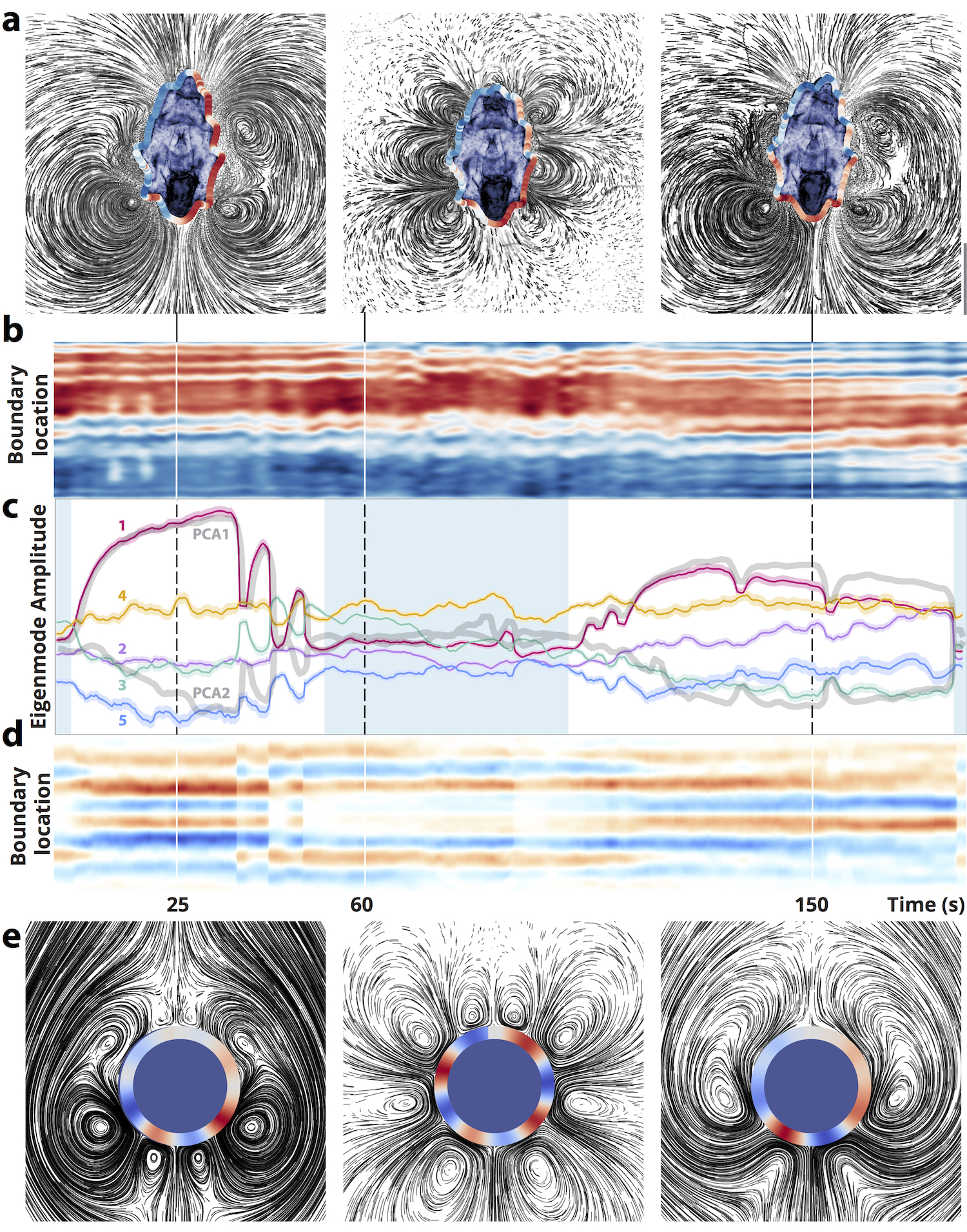}
\caption{
{\bf Time-dependent feeding versus swimming behavioral switches.} ({\bf A.}) The flow fields generated by an eight-week-old larva at three different timepoints in a three minute video. The tangential velocity component along the surface is overlaid as a color-coded outline on each image (Pathline length: 9 s, Scale bar: 500 $\mu$m). ({\bf B.}) A kymograph of the magnitude of the tangential velocity component along the swimmer surface at all timepoints, with the top of the animal in the previous images vertically centered (darkest red: $292\;\mu$m/s clockwise, white: zero, darkest blue: $340\;\mu$m/s counter-clockwise). ({\bf C.}) The relative amplitudes of the first five terms of a least-squares fit of the squirmer model to each frame of the video, with standard errors of the fit parameters indicated by shaded widths. The first-order "swimming" mode corresponds to the magenta trace. The amplitudes of data projections onto the first two principal components are underlaid in gray. Blue boxes underlaid indicate when swimming or feeding dominate in the original video. ({\bf D.}) A surface velocity kymograph for the best-fit squirmer models from the previous panel ({\bf E.}) Simulated flow fields for the best-fit squirmer model of the three movie frames shown in (A), with tangential velocity component color-coded on the squirmer boundary.}
\label{modes}
}
\end{figure}  

Next we investigate the temporal dynamics of the observed larval flow fields. Figure 4A shows three snapshots of the time-varying flow field around the organism, which changes dramatically over $\sim\! 10$ s timescales (Supplementary Video 11)---representing a known "arresting" behavior of starfish and other larvae in which short pauses punctuate regular swimming\cite{lacalli1990ciliary,strathmann2006good,mackie1969electrical} (Supplementary Videos 2, 3). In order to quantify these flow field changes, particle image velocimetry (PIV) is used to extract the velocity field as a function of time (Supplementary Methods) \cite{openpiv2010}. Kymographs of the surface velocity profile illustrate that the flow transiently switches direction in the regions between two defects (Figure 4B, 4D, boundaries between red/blue regions). This illustrates that the defects are dynamic, and it corroborates previous studies that report localized reversals of portions of the ciliary band\cite{lacalli1990ciliary,strathmann1971feeding,hart1991particle,mackie1969electrical}.

Principal component analysis (PCA) applied to the entire velocity field time series reveals two principal components corresponding to distinct far-field swimming and near-field feeding vortex behaviors. Projecting each frame of the PIV data onto the two principal components reveals that the animal switches back and forth between the two modes (Fig. 4C, gray traces), consistent with the transitions seen in freely-swimming animals (Supplementary Video 2)\cite{lacalli1990ciliary}. Direct fitting to the individual modes of the squirmer model (Fig. 4C, colored traces) confirms that the two anti-correlated principal components correspond to feeding and swimming: the first-order swimming mode (magenta trace) correlates with the first principal component, while the next four non-swimming modes resemble the anti-correlated component. These non-swimming modes lead to a higher capture cross section when particle simulations are performed on each frame of the time series of best-fit squirmer models (Supplementary Analysis). Plots of the velocity field corresponding to each best fit squirmer model (Fig. 4E) qualitatively resemble the original movie frames (Fig. 4A).

The correspondence of the PCA and squirmer fits suggests that the larval behavior is dominated by characteristic "eigenstrokes"---feeding and swimming---that can each be represented as amplitudes of a finite set of coefficients in the squirmer model. Low-dimensional behavior has been observed in other living systems\cite{berman2014mapping,stephens2008dimensionality,jordan2013behavioral}; however, in our system these characteristic eigenmodes can be directly linked with both the governing physical equations (Stokes' flow) and opposing biological functions (feeding and swimming). This behavioral switching may represent an adaptive strategy for heterogenous environments, analogous to the "roaming/dwelling" transition observed in other organisms that constitutes an optimal trade-off between global exploration and local exploitation \cite{berg1993random}.

The complex spatiotemporal properties of the starfish larval flow field extend earlier results suggesting that the feeding/swimming dichotomy may be a fundamental design principle for marine invertebrate larvae \cite{strathmann2006good,von2013pilidium,malkiel2003three}. Future field studies will further establish the relevance of the capture cross section for the heterogenous ocean environment, an open question in the literature on zooplankton grazing \cite{fenchel2002microbial,marrase1990grazing,rothschild1988small}. Additionally, in such a setting, vortex arrays may prove to have alternative functions, such as for masking mechanical strain fields that attract predators \cite{kiorboe2011zooplankton}. This would suggest that specific high-order power law decay terms in the squirmer model may be favored, constraining the placement and control of defects along the body.

Our work establishes a feeding/swimming tradeoff exemplified by larvae like echinoderms that use a single ciliary band to accomplish both functions. Future work will determine whether other planktonic larvae with different band morphologies (e.g. opposed-band feeders, such as gastropod larvae) experience similar hydrodynamical constraints\cite{von2013pilidium}. Importantly, single-band larvae rely on innervation to modulate separate functions within the band, leading some to speculate that the ciliary band preceded the evolution of the spinal cord\cite{lacalli1990ciliary,lowe2015deuterostome,nielsen1987structure}. Prior studies have indicated that the presence of $MgCl_{2}$ in solution can inhibit larval muscle control and ciliary reversal\cite{strathmann1971feeding}, causing the larvae to lose the ability to reverse direction after encountering obstacles (Supplementary Video 15). This suggests the need for further work using pharmacological techniques to relate the dynamics of the ciliary defects to the nervous system. This will reveal whether the observed ciliary defects implicate previously-identified neurociliary control mechanisms in invertebrate larvae,\cite{burke1983structure,lacalli1990ciliary} and whether their locations arise from the finite axonal contact range of neurons responsible for ciliary control \cite{jekely2011origin}.

Behavior, adaptation, and morphology in marine invertebrate larvae may be linked through the fluid mechanics of the ciliary band.\cite{strathmann1971feeding,hart1991particle,von2013pilidium} Our findings have additional potential implications for pathologies involving collective ciliary dyskinesia.\cite{marshall2008cilia} Moreover, it may also be possible to implement bioinspired "driven defect" boundary conditions in artificial systems as a means of stably generating complex flow structures\cite{shields2010biomimetic}, with applications to soft robotics and impingement filtration \cite{rubenstein1977mechanisms}.

\clearpage

\section{Contributions}

All authors designed and planned research collaboratively and performed experiments.  W.G. developed the theoretical model, and performed the numerical simulations and PIV analysis. All authors wrote the paper.

\section{Data Availability}

The data that support the plots within this paper and other findings of this study are available from the corresponding author upon request

\section{Acknowledgments}

We thank C. Lowe, D. N. Clarke, K. Uhlinger, and Stanford BIOS 236; also R. Konte,  L.-M. Joubert, M. S. Bull, D. Krishnamurthy, and the Prakash Lab. This work was supported by an ARO MURI Grant W911NF-15-1-0358 and an NSF CAREER Award (to M.P.) and an NSF GRFP DGE-114747 (to W.G.)

\bibliography{starfish_cites} 
\bibliographystyle{naturemag}

\end{document}